\documentclass[aps,pra,reprint,amsmath,amssymb,floatfix,superscriptaddress]{revtex4-2}
\usepackage{graphicx}
\usepackage{bm}
\usepackage{amsthm}
\usepackage{hyperref}

\newcommand{\heff}{\hbar_{\mathrm{eff}}}
\newcommand{\Lz}{\hat{L}_z}
\newcommand{\rTF}{\rho_{\mathrm{TF}}}
\newcommand{\RTF}{R_{\mathrm{TF}}}
\newcommand{\muloc}{\mu_{\mathrm{loc}}}
\newcommand{\bx}{\mathbf{x}}
\newcommand{\bv}{\mathbf{v}}
\newcommand{\ez}{\hat{\mathbf{z}}}
\newcommand{\dd}{\mathrm{d}}
\newcommand{\wK}{\omega_{K}}
\newcommand{\wcone}{\omega_{c1}}
\newcommand{\wLandau}{\omega_{\mathrm{L}}^{\mathrm{surf}}}
\newcommand{\wshed}{\omega_{\mathrm{shed}}}
\newcommand{\ETF}{\mathcal{E}_{\mathrm{TF}}}
\newtheorem{theorem}{Theorem}
\newtheorem{proposition}{Proposition}

\begin{document}

\title{Surface corrections and vortex energetics in the Thomas--Fermi limit\\
of a self-bound unitary Fermi superfluid}

\author{Bao-Duy Le}
\affiliation{Faculty of Physics and Engineering Physics, University of Science, Ho Chi Minh City, Vietnam}
\affiliation{Vietnam National University, Ho Chi Minh City, Vietnam}
\email[Email: ]{24130012@student.hcmus.edu.vn}
\begin{abstract}
We study a spin-balanced unitary Fermi superfluid bound by an attractive $1/r$
interaction within an extended Thomas--Fermi density functional. Its large-particle-number
limit is the Lane--Emden polytrope of index $3/2$. We prove the leading gradient
correction to the ground-state energy and strong convergence of the density amplitude.
A rescaled surface problem predicts the next energy correction and the leading
integrated density error. Matching a vortex core to the polytropic bulk gives the
energetic threshold of a singly quantized central vortex, including the constant
beyond the leading logarithm. Linearizing the same surface problem predicts a local energetic threshold
for collective surface excitations that, in the natural TF frequency units, decreases more slowly with
particle number than the central-vortex threshold. Radial stationary, spherical
Bogoliubov--Poisson, and axisymmetric real-time calculations test the predicted
energy corrections and vortex matching, and resolve finite-size surface-mode minima
whose approach to the continuum surface coefficient is followed to $N\simeq4\times10^6$.
For the reference gradient coefficient the vortex constant is $c_1=-0.535$, and the
surface Landau scale is $1.417\,\omega_K(\hbar_{\rm eff}^2/2R_{\rm TF})^{1/6}\propto N^{-1/9}$;
sector energies at $N\simeq4.6\times10^5$, computed after the fit, agree with the
extrapolated vortex threshold to $10^{-4}$. These results distinguish the accuracy
of the Thomas--Fermi bulk from the gradient physics that controls the surface and
vortex core, and separate vortex energetics from the open problem of nonlinear
vortex entry.

\end{abstract}

\maketitle

\section{Introduction}

In a self-bound superfluid both the density profile and the confining potential
respond to the fluid, so quantized circulation competes with a deformable surface
rather than with a prescribed trap. Vortices in strongly interacting Fermi gases
have been observed \cite{Zwierlein2005} and studied with superfluid density
functionals \cite{BulgacYu2003}, and rotating quantum droplets provide self-bound
vortex states \cite{Cidrim2018,Tengstrand2019}. We consider a spin-balanced
Fermi gas at unitarity \cite{Giorgini2008} with an attractive $1/r$ interaction
treated at the Hartree level, described by an orbital-free extended
Thomas--Fermi (ETF) functional \cite{SalasnichToigo2008,SalasnichSurface2010}.
The attraction is motivated by the light-induced interaction proposed for cold
atoms \cite{ODell2000} and becomes Newtonian gravity for $u=Gm^2$; its
Thomas--Fermi (TF) equilibrium is the Lane--Emden polytrope of index $3/2$
\cite{ChandrasekharStellar,AuchmutyBeals1971,LiebYau1987}. The same profile emerges
rigorously as the blow-up limit of neutron-star minimizers near the Chandrasekhar
limit, both in the Chandrasekhar model \cite{Nguyen2019Chandrasekhar} and in
Hartree--Fock--Bogoliubov theory, which includes pairing \cite{Nguyen2019HFB}.
Lopes \cite{Lopes2026} treats the Bohm term of a two-species self-gravitating
fermion model as a global surface-energy contribution to the mass--radius
relation; here we isolate its asymptotic role through a universal inner problem.
Self-gravitating Bose condensates have been studied as dark-matter halos,
including their equilibria \cite{Chavanis2011} and vortices
\cite{KainLing2010,RindlerDaller2012}; the paired Fermi fluid differs through its
$\rho^{5/3}$ equation of state, its compact polytropic surface, and its
circulation quantum $\pi\hbar/m$.
The small parameter is $\heff=N^{-1/3}/2$, with $N$ the number of fermions.
The gradient energy is suppressed in the bulk as $N^{-2/3}$ but remains essential
near the free surface and a vortex core. For trapped Bose gases, vortex energies,
corrections beyond TF, and surface criteria for vortex entry are well studied
\cite{Lundh1997,FetterSvidzinsky2001,Fetter2009,PitaevskiiStringari}, including
rigorous TF limits under fast rotation \cite{NguyenRougerie2022} and vortex
patterns near the critical rotation speed \cite{LeNguyen2026}. Lundh, Pethick,
and Smith treated the TF boundary layer of a trapped Bose gas \cite{Lundh1997},
and Anglin obtained a local Landau criterion from the surface spectrum in a linear
potential \cite{Anglin2001}. We carry these constructions over to a $\rho^{5/3}$
energy density in a self-generated potential and match the vortex core to the
Lane--Emden geometry. The new quantitative results are the static surface
coefficients, the constant term in the central-vortex threshold, and the surface
Landau scale. Microscopic descriptions of the unitary gas treat vortices and
interfaces separately \cite{Kryjevski2008}, and large-charge calculations
distinguish the bulk gradient expansion from an edge problem \cite{Hellerman2024};
our ``universal'' coefficients refer to parameter-free inner equations within the
ETF model, not to a microscopic determination.
Theorem~\ref{thm:energy} and Proposition~\ref{prop:exact} are exact statements
within the functional; Propositions~\ref{prop:layer}, \ref{prop:wc1}, and
\ref{prop:nuc} are formal asymptotic results, without remainder bounds, that we
test numerically. Nonlinear vortex entry and rotating few-vortex and lattice
equilibria remain open (Sec.~\ref{sec:discussion}).

\section{Model}
\label{sec:model}

\subsection{Energy functional}

We consider $N$ fermions of mass $m$ in two spin states with equal populations,
interacting through an $s$-wave contact interaction at unitarity and a long-range
attraction $-u/|\mathbf r-\mathbf r'|$. At zero temperature the superfluid is described
by an effective pair field $\Psi$, with total density $n=2|\Psi|^2$ and energy
\cite{SalasnichToigo2008,SalasnichSurface2010}
\begin{align}
E[\Psi]=\int_{\mathbb R^3}\Big[&\frac{\hbar^2}{4m}|\nabla\Psi|^2
+\frac35\,\xi_B\,\frac{\hbar^2}{2m}(3\pi^2)^{2/3}n^{5/3}\nonumber\\
&+\frac12\,n\,\Phi
+(\lambda-\tfrac14)\frac{\hbar^2}{8m}\frac{|\nabla n|^2}{n}\Big]\,\dd\mathbf r,
\label{eq:E}
\end{align}
with $\xi_B$ the Bertsch parameter \cite{Ku2012} and
$\Phi(\mathbf r)=-u\int_{\mathbb R^3}n(\mathbf r')\,|\mathbf r-\mathbf r'|^{-1}\,\dd\mathbf r'$, $\mathbf r\in\mathbb R^3$.
The long-range interaction enters only through this Hartree potential; changes
to the short-range unitary equation of state are outside the model.
Writing $\Psi=\sqrt{n/2}\,e^{i\theta}$ separates the flow energy
$\tfrac12mn|\bv|^2$, with $\bv=\hbar\nabla\theta/(2m)$, from the amplitude-gradient
energy $\lambda\hbar^2|\nabla n|^2/(8mn)$. The circulation quantum $\pi\hbar/m$
is fixed by pairing. The coefficient $\lambda$ is an independent parameter of the
density functional; the pair mass does not determine it.

We choose $\lambda=1/4$ as the reference model, for which the last term in
Eq.~(\ref{eq:E}) vanishes and the dynamics has the simplest pair-field
Schr\"odinger form. This value has been used in ETF descriptions, but is not the
Monte Carlo best-fit coefficient of Ref.~\cite{SalasnichToigo2008}: its erratum
gives $\xi_B=0.468$ and $\lambda=0.086$ for that fit
\cite{SalasnichErratum2010}. We instead use the measured bulk value
$\xi_B=0.37$ \cite{Ku2012} and treat $\lambda=1/4$ as a specified gradient
model, without claiming a joint microscopic calibration. General $\lambda>0$
is retained in the static rescaling and core problem; calculations at
$\lambda=1/9$ test the sensitivity of the vortex threshold. That latter value is
the Kirzhnits coefficient for an ideal Fermi gas \cite{Kirzhnits1957}, not a
calibration of the interacting unitary gas.

\subsection{Scaled units}

The attraction defines $a_u=\hbar^2/(mu)$. The TF balance between
$\muloc(n)=\xi_B\hbar^2(3\pi^2n)^{2/3}/2m$ and $\Phi$ fixes the radius
$r_N=a_uN^{-1/3}$, the energy per particle $E_N=uN/r_N$ and the frequency
$\Omega_N=(E_N/mr_N^2)^{1/2}$. In the variables $\bx=\mathbf r/r_N$, $\rho=nr_N^3/N$
($\int_{\mathbb R^3}\rho\,\dd\bx=1$), $\psi=\sqrt\rho\,e^{i\theta}$, $t\to\Omega_Nt$, $\omega=\Omega/\Omega_N$,
and energies per particle in units of $E_N$, the dynamics at $\lambda=1/4$ in a frame rotating about
$\ez$ is
\begin{gather}
i\heff\partial_t\psi=\Big[-\frac{\heff^{2}}{2}\nabla^2+\kappa|\psi|^{4/3}+\Phi
-\omega\heff\Lz\Big]\psi,\label{eq:GP}\\
\nabla^2\Phi=4\pi|\psi|^2,\nonumber
\end{gather}
with $\Lz=-i(x\partial_y-y\partial_x)$ and
\begin{equation}
\heff=\frac{\hbar}{2mr_N^2\Omega_N}=\frac12N^{-1/3},\qquad
\kappa=\frac{\xi_B}{2}(3\pi^2)^{2/3}.
\label{eq:params}
\end{equation}
All frequency scalings with $N$ quoted below refer to $\omega=\Omega/\Omega_N$.
At fixed $u$ and $m$, $\Omega_N\propto N$, so a decreasing dimensionless
threshold need not be a decreasing laboratory frequency.
The energy in the rotating frame is
\begin{align}
\mathcal E_\omega[\psi]=\int_{\mathbb R^3}\Big[&\frac{\heff^2}{2}|\nabla\psi|^2
+2\Big(\lambda-\frac14\Big)\heff^2\big|\nabla|\psi|\big|^2\nonumber\\
&\hspace{-2em}+\frac35\kappa|\psi|^{10/3}+\frac12|\psi|^2\Phi-\omega\heff\psi^*\Lz\psi\Big]\dd\bx.
\label{eq:Escaled}
\end{align}
For a real nonnegative $\psi=\sqrt\rho$ the two gradient terms combine to
$2\lambda\heff^2|\nabla\sqrt\rho|^2$, so static results at any $\lambda$ follow from
those at $\lambda=1/4$ with $\heff\to2\sqrt\lambda\,\heff$. Rescaling
$\bx\to\kappa\bx$ removes $\kappa$ at the cost of $\heff\to\heff/\sqrt\kappa$ and
$\omega\to\omega\kappa^{3/2}$, so its dependence can be reconstructed from this rescaling; we use $\xi_B=0.37$
\cite{Ku2012}, $\kappa=1.7706$. The gradient, flow and $\rho^{5/3}$ terms scale as
(length)$^{-2}$ at fixed norm and $W=\tfrac12\int_{\mathbb R^3}\rho\Phi\,\dd\bx$ as (length)$^{-1}$, so
stationary states obey $2(T+E_{\rm int})+W=0$, with $T$ the kinetic and $E_{\rm int}$
the $\rho^{5/3}$ energy.

\paragraph*{Notation.} Spatial integrals are written $\int_{\mathbb R^3}f(\bx)\,\dd\bx$,
with $\bx\in\mathbb R^3$ and $\dd\bx$ the Lebesgue measure; $r=|\bx|$, and $(s,\varphi,z)$
are cylindrical coordinates about $\ez$. Function spaces are taken on $\mathbb R^3$,
$L^p=L^p(\mathbb R^3)$ and $H^1=H^1(\mathbb R^3)$, and $\|\cdot\|_p$ is the norm of $L^p$.

\section{Semiclassical analysis of the static star}
\label{sec:static_theory}

At $\omega=0$ the minimizer may be taken real and nonnegative, since
$|\nabla|\psi||\le|\nabla\psi|$, and Eq.~(\ref{eq:Escaled}) becomes
\begin{gather}
E_{\heff}[\rho]=\heff^2\,\mathcal T[\rho]+\ETF[\rho],\label{eq:split}\\
\mathcal T[\rho]=2\lambda\int_{\mathbb R^3}|\nabla\sqrt\rho|^2\,\dd\bx,\nonumber
\end{gather}
\begin{equation}
\begin{split}
\ETF[\rho]={}&\frac35\kappa\int_{\mathbb R^3}\rho^{5/3}\,\dd\bx\\
&-\frac12\int_{\mathbb R^3}\!\int_{\mathbb R^3}\frac{\rho(\bx)\rho(\bx')}{|\bx-\bx'|}\,\dd\bx\,\dd\bx'.
\end{split}
\label{eq:ETF}
\end{equation}
This is an attractive Hartree counterpart of Thomas--Fermi--von Weizs\"acker
theory \cite{BenguriaBrezisLieb1981}, with the polytropic TF limit related to
gravitational fermion models \cite{LiebYau1987}. We set $\lambda=1/4$ below.

\subsection{TF limit}

The minimizer of $\ETF$ at unit mass exists and is unique up to translations
\cite{AuchmutyBeals1971,LiebYau1987,Nguyen2019Chandrasekhar}: $\rTF=\rho_c\vartheta^{3/2}(r/\alpha)$ with
$\vartheta$ the Lane--Emden function of index $3/2$, first zero $\zeta_1=3.65375$,
$\varpi=-\zeta_1^2\vartheta'(\zeta_1)=2.71406$, $\alpha=\RTF/\zeta_1$, and
\begin{gather}
\RTF=\frac{\zeta_1\varpi^{1/3}}{(4\pi)^{2/3}}\,\kappa,\qquad
\rho_c=\frac{4\pi}{\varpi^2\kappa^3},\nonumber\\
\mu_{\rm TF}=-\frac{1}{\RTF},\qquad
E_{\rm TF}=-\frac{3}{7\RTF}.
\label{eq:TF0}
\end{gather}
For $\xi_B=0.37$, $\RTF=1.66952$, $\rho_c=0.307337$, $\mu_{\rm TF}=-0.598975$,
$E_{\rm TF}=-0.256704$; we define $\wK\equiv\RTF^{-3/2}=0.463568$, the Kepler frequency
at the TF radius. The central healing length is
$\xi_0=\heff/(\kappa\rho_c^{2/3})^{1/2}=1.114\,\heff$, of the order of the
interparticle spacing.

\begin{theorem}
\label{thm:energy}
Let $E(\heff)=\inf\{E_{\heff}[\rho]:\rho\ge0,\ \sqrt\rho\in H^1(\mathbb R^3),\ \int_{\mathbb R^3}\rho\,\dd\bx=1\}$ and
$A=\mathcal T[\rTF]=\tfrac12\int_{\mathbb R^3}|\nabla\sqrt{\rTF}|^2\,\dd\bx$. Then
\begin{enumerate}
\item[(i)] $0\le E(\heff)-E_{\rm TF}\le A\,\heff^2$ for every $\heff>0$;
\item[(ii)] $\lim_{\heff\to0}[E(\heff)-E_{\rm TF}]/\heff^2=A$;
\item[(iii)] if $\rho_{\heff}$ is a minimizer, then up to translations the amplitude
converges, $\sqrt{\rho_{\heff}}\to\sqrt{\rTF}$ strongly in $H^1(\mathbb R^3)$; the density
converges in $L^p(\mathbb R^3)$, $1\le p\le3$; and the Lagrange multiplier $\mu(\heff)$ of the normalization constraint satisfies $\mu(\heff)-\mu_{\rm TF}=\tfrac53A\heff^2+o(\heff^2)$.
\end{enumerate}
The natural variable is the amplitude $\sqrt\rho\in H^1$; no statement about $\rho$ in
$H^1$ is made or needed.
For $\xi_B=0.37$, $A=1.90474$.
\end{theorem}

\begin{proof}
$A$ is finite: near the edge $\rTF\propto(\RTF-r)^{3/2}$, so
$|\nabla\sqrt{\rTF}|^2\propto(\RTF-r)^{-1/2}$ is integrable. 
(i) The upper bound follows from the trial density $\rTF$. For the lower bound, $\mathcal T\ge0$ gives
$E_{\heff}[\rho]\ge\ETF[\rho]\ge E_{\rm TF}$. (ii) Let $\rho_{\heff}$ be any admissible density with
	$E_{\heff}[\rho_{\heff}]\le E(\heff)+\heff^3$ (a minimizer, when it exists). Write
	\begin{equation}
		E_{\heff}[\rho_{\heff}]-E_{\rm TF}=\heff^2\mathcal T[\rho_{\heff}]+(\ETF[\rho_{\heff}]-E_{\rm TF})
	\end{equation}
	with both terms nonnegative. By (i), $\mathcal T[\rho_{\heff}]\le A+\heff$ and
	$\ETF[\rho_{\heff}]-E_{\rm TF}\le A\heff^2+\heff^3$, so $\rho_{\heff}$ is a minimizing sequence for
	$\ETF$ and $\sqrt{\rho_{\heff}}$ is bounded in $H^1$. The TF energy at mass $M$ scales as
	$M^{7/3}E_{\rm TF}$, which gives strict subadditivity because $E_{\rm TF}<0$. The negative
	infimum excludes vanishing, and strict subadditivity excludes dichotomy in the
	concentration-compactness argument \cite{Lions1984}. Tightness up to translations,
	combined with the local compactness of the bounded $H^1$ amplitudes, gives
	strong $L^1$ convergence of the densities. Interpolation gives convergence in
	$L^{6/5}$, hence continuity of the Hartree term; lower semicontinuity of the
	local term identifies the limit as a TF minimizer. Uniqueness fixes it to $\rTF$. Weak lower semicontinuity gives
	$\liminf\mathcal T[\rho_{\heff}]\ge\mathcal T[\rTF]=A$. Combined with
	$(E-E_{\rm TF})/\heff^2\ge\mathcal T[\rho_{\heff}]-\heff$ and the upper bound, this proves (ii)
	and $\mathcal T[\rho_{\heff}]\to A$. (iii) Let $w_{\heff}=\sqrt{\rho_{\heff}}$ and
	$w=\sqrt{\rTF}$. Concentration compactness gives $\rho_{\heff}\to\rTF$ in $L^1$ (after
	translation), and $\|w_{\heff}-w\|_2^2\le\|\rho_{\heff}-\rTF\|_1$ because
	$(\sqrt a-\sqrt b)^2\le|a-b|$; hence $w_{\heff}\to w$ in $L^2$. Since $w_{\heff}$ is bounded in
	$H^1$, $w_{\heff}\rightharpoonup w$ weakly in $H^1$, and
	$\|\nabla w_{\heff}\|_2^2=2\mathcal T[\rho_{\heff}]\to2A=\|\nabla w\|_2^2$; weak convergence with
	convergence of the norm is strong convergence in $\dot H^1$, and with the $L^2$ limit, in
	$H^1$. By Sobolev, $w_{\heff}\to w$ in $L^q$ for $2\le q\le6$, and Hölder's inequality applied to
	$\rho_{\heff}-\rTF=(w_{\heff}-w)(w_{\heff}+w)$ gives convergence of $\rho_{\heff}$ in $L^p$,
	$1\le p\le3$. Finally, from the virial identity, $E=-(T+E_{\rm int})$ and
	$\mu=-3T-\tfrac73E_{\rm int}$, so
	$\mu-\mu_{\rm TF}=\tfrac73(E-E_{\rm TF})-\tfrac23\heff^2\mathcal T[\rho_{\heff}]
	=\tfrac53A\heff^2+o(\heff^2)$.
 
\end{proof}

Theorem~\ref{thm:energy} states precisely what the Bohm term does in the TF limit: to
leading order it contributes its own expectation value in the TF state,
$A\heff^2$, and nothing else. All other effects of the gradient term are of
higher order.

\subsection{Surface boundary layer}

The next correction comes from the edge, where the derivative of the TF amplitude is
singular. Near $r=\RTF$, $\Phi-\mu\simeq F(r-\RTF)$
with $F=\RTF^{-2}$. With $r-\RTF=\delta\eta$ and $\sqrt\rho=(F\delta/\kappa)^{3/4}\chi_*(\eta)$,
the gradient, local and potential terms balance for
\begin{equation}
\delta=\Big(\frac{\heff^2}{2F}\Big)^{1/3}=\Big(\frac{\heff^2\RTF^2}{2}\Big)^{1/3},
\label{eq:delta}
\end{equation}
and $\chi_*$ solves the universal, parameter-free problem
\begin{equation}
\begin{gathered}
-\chi_*''+\big(\eta+\chi_*^{4/3}\big)\chi_*=0,\\
\chi_*\to(-\eta)^{3/4}\ (\eta\to-\infty),\quad \chi_*\to0\ (\eta\to\infty).
\end{gathered}
\label{eq:inner}
\end{equation}
Matching to the outer TF solution gives the following.

\begin{proposition}[formal]
\label{prop:layer}
Assume that $\rho_{\heff}$ is described by the outer TF solution plus $O(\heff^2)$ in the
bulk and by Eq.~(\ref{eq:inner}) in the layer, with the two matching in an overlap region
$\delta\ll\RTF-r\ll\RTF$. Then, to leading order in matched asymptotics,
\begin{gather}
E(\heff)=E_{\rm TF}+A\heff^2-B\heff^{7/3}+o(\heff^{7/3}),\label{eq:Eexp}\\
\|\rho_{\heff}-\rTF\|_1=P_\epsilon\heff^{5/3}\,[1+o(1)],\nonumber
\end{gather}
with $B=-4\pi\RTF^2F^{5/2}\kappa^{-3/2}(2F)^{-7/6}K$ and
$P_\epsilon=4\pi\RTF^2(F/\kappa)^{3/2}(2F)^{-5/6}M_1$. The surface constants are defined by
\begin{align}
K&=\int_{\mathbb R}\left[\mathcal G(\chi_*)-\mathcal G(\chi_{\rm TF})\right]\dd\eta,
\nonumber\\
M_1&=\int_{\mathbb R}|\chi_*^2-\chi_{\rm TF}^2|\,\dd\eta,
\nonumber
\end{align}
where $\chi_{\rm TF}=(-\eta)_+^{3/4}$ and
$\mathcal G(f)=f'^2+\tfrac35 f^{10/3}+\eta f^2$. Solving Eq.~(\ref{eq:inner}) gives
$K=-0.66225$ and $M_1=0.77359$, hence $B=1.1180$ and $P_\epsilon=3.2589$.
\end{proposition}

The constants are computed by solving Eq.~(\ref{eq:inner}) as a boundary-value problem on
$[-\eta_1,\eta_2]$ with $\chi_*(-\eta_1)$ from the asymptotic series and adding the analytic tails.
For $\eta\to-\infty$, $\chi_*^{4/3}=-\eta-\tfrac{3}{16}\eta^{-2}+\dots$, so
$\chi_*^2-(-\eta)^{3/2}\simeq-\tfrac{9}{32}(-\eta)^{-3/2}$ and the tail of $M_1$ beyond
$-\eta_1$ is $\tfrac{9}{16}\eta_1^{-1/2}$; it is not negligible ($0.073$ at $\eta_1=60$). With the
tails included, $K$, $B$ and $M_1$ change by less than $2\times10^{-5}$, $3\times10^{-5}$
and $3\times10^{-5}$ between $\eta_1=40$, $60$ and $80$ and between $\eta_2=12$ and $18$.

The surface layer lowers the energy below the first-order estimate because the
density relaxes into the classically forbidden region; the gain scales as
(layer volume)$\times$(energy density), $\delta\cdot\delta^{5/2}\propto\heff^{7/3}$.
The same layer redistributes a mass $\propto\delta^{5/2}\propto\heff^{5/3}$, which
dominates the $L^1$ error because it is larger than the bulk $O(\heff^2)$ change of
the profile. The signed leading layer mass vanishes, so the integrated error does not
require a bulk normalization change of order $\heff^{5/3}$. The cancellation
and the energy subtraction are detailed in Appendix~\ref{app:surface}.
Both coefficients are tested in Sec.~\ref{sec:static}.

\section{Rotation}
\label{sec:rotation}

\subsection{Exact reduction for a central vortex}

\begin{proposition}
\label{prop:exact}
For normalized axisymmetric states $\psi=f(s,z)e^{i\ell\varphi}$,
$\mathcal E_\omega[\psi]=E_{\rm lab}[f,\ell]-\omega\heff\ell$.
The profiles minimizing the laboratory energy within each sector are therefore
independent of $\omega$. The $\ell=1$ and $\ell=0$ sector energies cross at
\begin{equation}
\wcone=\frac{E_1-E_0}{\heff},\qquad E_\ell=\min_f E_{\rm lab}[f,\ell].
\label{eq:wc1exact}
\end{equation}
\end{proposition}

The rotation term is constant in each sector, which proves the reduction. The
$\ell=1$ state carries angular momentum $\heff$ per particle, and $\wcone$ is
the secant slope between the two sector energies as functions of angular momentum.
It is an energetic comparison of a central vortex with the vortex-free state;
it does not establish equivalence of the fixed-$L$ and fixed-$\omega$ ensembles
or global minimization over all rotating configurations. Indeed,
$E-\omega L$ is unbounded below in free space for nonzero $\omega$.
We consider intrinsic rotation about the center of mass, with translations and
center-of-mass motion removed.

Other sectors cross $\ell=0$ at
$\omega_\ell=(E_\ell-E_0)/(\ell\heff)$. The matching below gives
\begin{equation}
	\omega_\ell\RTF^2/(C_{3/2}\heff)=\ell(X+c_{\rm geo})+\gamma_\ell/\ell
\end{equation}
for fixed $\ell$, where $X=\ln(\RTF/\xi_0)$ and the core equation contains
$\ell^2g/x^2$. Thus $\omega_\ell/\wcone\to\ell$ as $\heff\to0$.
Direct sector calculations give $\omega_2/\wcone=1.294$, $1.445$, $1.548$ and
$\omega_3/\wcone=1.432$, $1.746$, $1.977$ at $\heff=0.05$, $0.025$, $0.0125$.
These comparisons support the singly quantized central vortex as the lowest
threshold among the sectors tested. Curved, off-center, and multiple separated
vortices require a broader variational comparison and are not excluded here.

\subsection{Asymptotics of $\wcone$ and the core constant}
\label{sec:wc1asym}

For $\heff\to0$, $E_1-E_0$ is computed by matched asymptotics with three regions:
the core ($s\lesssim\xi$), the bulk ($\xi\ll s<R_\perp(z)$) and the surface layer. In the
bulk the velocity is exactly $\heff/s$ (quantized circulation, axial symmetry) and the
density approaches $\rTF$ away from the core and surface. At fixed total mass, the linear bulk energy cost of restoring the core
depletion cancels the chemical-potential subtraction in the local core problem.
The remaining bulk relaxation does not affect the displayed
$O(\heff^2)$ vortex-energy coefficient under the matching assumptions
(Appendix~\ref{app:vortex}). In the core the local density is
$\rTF(0,z)$ and the local healing length is $\xi(z)=\heff/[\kappa\rTF(0,z)^{2/3}]^{1/2}$.
With $g(x)=|\psi|/\sqrt{\rho_\infty}$ and $x=s/\xi$, the core profile solves
\begin{equation}
\begin{gathered}
-4\lambda\Big(g''+\frac{g'}{x}\Big)+\frac{g}{x^2}+2\big(g^{7/3}-g\big)=0,\\
g(0)=0,\quad g(\infty)=1,
\end{gathered}
\label{eq:core}
\end{equation}
and a straight vortex in a uniform fluid has energy per unit length
$\pi\rho_\infty\heff^2[\ln(L_\perp/\xi)+\gamma(\lambda)]$, with $L_\perp\gg\xi$ its transverse size and
\begin{equation}
\gamma(\lambda)=\lim_{\Lambda\to\infty}\Big\{\int_0^\Lambda\Big[4\lambda g'^2+\frac{g^2}{x^2}+2U(g)\Big]x\,\dd x-\ln\Lambda\Big\},
\label{eq:gamma}
\end{equation}
where $U(g)=\tfrac35(g^{10/3}-1)-(g^2-1)$. Integrating the bulk flow energy
density $\tfrac12\heff^2\rho/s^2$ along the vortex line and matching to the core gives the
following.

\begin{proposition}[formal]
\label{prop:wc1}
To leading order in $\heff$, including the constant term, and assuming the three-region
structure described above,
\begin{gather}
\wcone=C_{3/2}\,\frac{\heff}{\RTF^2}\Big[\ln\frac{\RTF}{\xi_0}+c_1\Big]+o(\heff),\label{eq:wc1}\\
c_1=c_{\rm geo}+\gamma(\lambda),\nonumber\\
C_{3/2}=\frac{\zeta_1^2\int_0^{\zeta_1}\vartheta^{3/2}(\zeta)\,\dd\zeta}{2\int_0^{\zeta_1}\vartheta^{3/2}(\zeta)\,\zeta^2\,\dd\zeta}=4.150,
\nonumber
\end{gather}
with
\begin{align}
c_{\rm geo}=\Big\langle&\ln\frac{R_\perp(z)}{\RTF}+\frac13\ln\frac{\rTF(0,z)}{\rho_c}\nonumber\\
&+\int_0^{R_\perp(z)}\Big[\frac{\rTF(s,z)}{\rTF(0,z)}-1\Big]\frac{\dd s}{s}\Big\rangle_z,
\label{eq:cgeo}
\end{align}
where $R_\perp(z)=(\RTF^2-z^2)^{1/2}$ and
\begin{equation}
	\langle f\rangle_z= \frac{\int_{-\RTF}^{\RTF}f(z)\rTF(0,z)\,\dd z}{\int_{-\RTF}^{\RTF}\rTF(0,z)\,\dd z} 
\end{equation}
is the average along the axis with weight $\rTF(0,z)$.
\end{proposition}

The three terms of $c_{\rm geo}$ account, respectively, for the finite transverse size of
each slice, the variation of the healing length along the axis, and the radial decrease
of the density that carries the $1/s$ flow. For the Lane--Emden profile
$c_{\rm geo}=-1.10355$. The core constant follows from Eq.~(\ref{eq:core}); we checked the
procedure on the standard Gross--Pitaevskii vortex, for which it gives
$0.72746$ against the known $\ln(1.464\sqrt2)=0.72775$ \cite{FetterSvidzinsky2001}.
Table~\ref{tab:gamma} lists $\gamma(\lambda)$; for $\lambda=1/4$,
$c_1=-1.10355+0.56845=-0.53509$.

The matching computes the energy of a central vortex under the assumed
core--bulk--surface structure. A trial-field construction motivates the upper
estimate, but controlling its remainder uniformly through the two vortex ends
would require an additional argument. A matching lower bound would also require
estimates analogous to those developed for trapped condensates
\cite{IgnatMillot2006,Aftalion2006}. We claim neither bound here; the formula is
tested against the axisymmetric sector energies in Sec.~\ref{sec:results}.

\paragraph*{Approach to the limit.} To compare Eq.~(\ref{eq:wc1}) with data at finite
$\heff$ one needs the form of the leading correction, not only the limit. Write
$y=\wcone\RTF^2/(C_{3/2}\heff)$ and $X=\ln(\RTF/\xi_0)$, so that $y-X\to c_1$. Two effects
are neglected in Proposition~\ref{prop:wc1}. (i) The outer flow integral uses $\rTF$
instead of the true density $\rho_{\heff}$. The effect of the background profile can be isolated approximately by evaluating the slice
formula of Appendix~\ref{app:vortex} on $\rho_{\heff}$,
\begin{align}
\omega_{\rm loc}(\heff)=\;&\pi\heff\int_{\mathbb R}\Big\{\rho_{\heff}(0,z)\big[\gamma-\ln\xi(z)\big]\nonumber\\
&+\!\int_0^\infty\!\big[\rho_{\heff}(s,z)\nonumber\\
&\quad-\rho_{\heff}(0,z)\mathbf 1_{[0,1]}(s)\big]\frac{\dd s}{s}\Big\}\dd z,
\label{eq:wloc}
\end{align}
with $\xi(z)$ from $\rho_{\heff}(0,z)$ and $\mathbf 1_{[0,1]}$ the indicator function of $[0,1]$; $\omega_{\rm loc}$ tends to Eq.~(\ref{eq:wc1}) as
$\heff\to0$. (ii) Gradient corrections to the energy per unit length of the line. One contribution is the axial amplitude-gradient energy, whose change contains
\begin{align}
	\tfrac12\heff^2\int_0^{R_\perp}(g^2-1)|\partial_z\sqrt\rho|^2\,2\pi s\,\dd s\nonumber\\
	\simeq-\tfrac{3\pi}{4}\heff^2\xi^2|\partial_z\sqrt\rho|^2\ln(R_\perp/\xi),
\end{align}
since
$1-g^2\simeq\tfrac34x^{-2}$. In the bulk it is $O(\heff^4\ln\heff^{-1})$. Near the two
ends of the line, at a distance $d=\RTF-|z|$ from the pole, $\rho(0,z)\propto d^{3/2}$,
$\xi\propto\heff d^{-1/2}$ and $R_\perp\propto d^{1/2}$, so the integrand grows as
$\heff^4d^{-3/2}\ln(d/\heff)$ and is cut off where the slice description fails, at
$d\simeq\delta\propto\heff^{2/3}$. The result is $\heff^4\delta^{-1/2}[\ln(\delta/\heff)+O(1)]$,
which with $\delta/\heff\propto\heff^{-1/3}$ gives
\begin{equation}
y-X\simeq c_1+\heff^{5/3}\big(a\ln\heff^{-1}+b\big).
\label{eq:c1corr}
\end{equation}
Equation~(\ref{eq:c1corr}) is an extrapolation ansatz motivated by this
endpoint contribution, not a derived expansion with a controlled remainder.
The coefficients $a,b$ require a matched solution of the vortex ends.
Normalization and bulk relaxation must also be included in a complete
next-order calculation: the $1/s^2$ depletion can give a mass deficit of order
$\heff^2\ln\heff^{-1}$, whose quadratic relaxation may produce additional
logarithms in the energy. We therefore test sensitivity to both
$\heff^2\ln\heff^{-1}$ and $\heff^2\ln^2\heff^{-1}$ corrections, without claiming
that either exhausts the next-order terms.

\begin{table}[!tbp]
\caption{Core constant $\gamma(\lambda)$, Eq.~(\ref{eq:gamma}), and
$c_1=c_{\rm geo}+\gamma$ with $c_{\rm geo}=-1.10355$. ``No Bohm'' is the hydrodynamic TF
vortex, $g^{4/3}=[1-1/(2x^2)]_+$, without the amplitude-gradient term.}
\label{tab:gamma}
\begin{ruledtabular}
\begin{tabular}{lcc}
 & $\gamma$ & $c_1$\\ \hline
no Bohm term & 0.20639 & $-0.89716$\\
$\lambda=1/9$ (Kirzhnits) & 0.40096 & $-0.70259$\\
$\lambda=1/4$ (reference) & 0.56845 & $-0.53509$\\
$\lambda=0.4$ & 0.71957 & $-0.38397$\\
\end{tabular}
\end{ruledtabular}
\end{table}

\paragraph*{Role of the Bohm term in the vortex.} Without the amplitude-gradient term
the vortex still has a finite core, set by the centrifugal barrier, and
$\gamma_{\rm TF}=0.20639$. The Bohm term smooths the core and raises $\gamma$ by
$0.36207$. At $N=6.4\times10^4$, the matched formula evaluated on the same TF
background gives a $9.3\%$ increase relative to the value without the amplitude
gradient (the difference is $8.5\%$ of the reference value). This is an
asymptotic comparison; finite-$\heff$ equilibrium profiles also depend on $\lambda$. Because $\wcone$ is
$O(\heff\ln\heff^{-1})$, the vortex energy $\heff\wcone$ exceeds the static Bohm
correction $A\heff^2$ by a factor $\ln\heff^{-1}$: a single vortex is a larger
departure from TF than the gradient correction of the whole star.

\subsection{Surface spectrum and Landau threshold}
\label{sec:nuc}

For the spherical vortex-free minimizer, the rotating-frame frequency of a mode
with angular indices $(l,m_z)$ is exactly
$\omega_{l,m_z}(\omega)=\omega_l(0)-m_z\omega$.
The profile is unchanged by the rotation term, and its Bogoliubov operator
commutes with $\Lz$. Modes with $m_z=0$ and $m_z=l$ are degenerate in the
laboratory frame, so axisymmetric linear response determines the frequencies
needed for the co-rotating surface branch. We define its Landau threshold by
\begin{equation}
\wLandau=\min_{(l,\omega_l)\in\mathcal S_{\rm surf}}\frac{\omega_l}{l},\qquad l\ge2.
\label{eq:wnucdef}
\end{equation}
Here $\omega_l$ denotes the lowest surface ($f$-)mode at each $l$; translation
at $l=1$ is removed by fixing the center of mass. The set
$\mathcal S_{\rm surf}$ is the collective branch connected to the nodeless
hydrodynamic $f$-mode, with perturbations localized near the star. In the
surface limit this means $k\delta=O(1)$ and localization in
$|r-\RTF|=O(\delta)$. At finite $\heff$ we identify this branch by its radial
structure and continuity across successive $l$, and check localization using
both the density perturbation and the quasiparticle amplitudes
(Appendix~\ref{app:spectrum}). This prescription selects a collective response;
it is not a minimization over all positive eigenvalues at arbitrarily large $l$. It excludes spatially extended
particle-escape excitations and is not the infimum over the entire free-space
rotating spectrum, for which $E-\omega L$ is unbounded below.
Equation~(\ref{eq:wnucdef})
is a criterion for negative surface-excitation energy in the rotating frame.
The real-frequency shift alone does not imply a dynamical instability or
vortex formation. Coupling to a nonaxisymmetric drive, dissipation, and the
subsequent nonlinear evolution determine whether vortices enter.

At fixed $l$ in the TF limit, $\omega_l$ is the $f$-mode of the $3/2$ polytrope.
For large $l$ it approaches the surface gravity wave
$\omega_l^2\simeq Fk$, $k\simeq l/\RTF$, so
$\omega_l/l\simeq\wK/\sqrt l$. The classical decrease is regularized when
$k\delta=O(1)$ and the density-gradient term enters the surface dispersion.
Following the local construction of Ref.~\cite{Anglin2001}, we linearize the
surface problem for the unitary equation of state.

\begin{proposition}[formal]
\label{prop:nuc}
Let $q=k\delta$ and linearize the inner equation,
$i\partial_\tau\chi=[-\nabla^2+\eta+|\chi|^{4/3}]\chi$ (lengths in units of $\delta$, time
in units of $\heff/F\delta$), about $\chi_*$ with transverse wave number $q$. The
Bogoliubov frequencies $\nu$ satisfy
\begin{equation}
\nu^2 f=L_q\Big(L_q+\tfrac43\chi_*^{4/3}\Big)f,\quad
L_q=-\partial_\eta^2+q^2+\eta+\chi_*^{4/3},
\label{eq:bdg}
\end{equation}
whose lowest eigenvalue $\nu(q)$ obeys $\nu\to\sqrt{2q}$ for $q\to0$ (the gravity wave)
and $\nu\simeq q^2$ for $q\to\infty$. To leading order in $\delta/\RTF$, with the Cowling
approximation and $k=\sqrt{l(l+1)}/\RTF$,
\begin{equation}
\frac{\omega_l}{l}=\wK\Big(\frac{\heff^2}{2\RTF}\Big)^{1/6}\frac{\nu(q)}{\sqrt2\,q}\,
\frac{\sqrt{l(l+1)}}{l},
\label{eq:wl_layer}
\end{equation}
and
\begin{equation}
\wLandau=m_*\,\wK\Big(\frac{\heff^2}{2\RTF}\Big)^{1/6},\qquad l_*=q_*\frac{\RTF}{\delta},
\label{eq:wnuc}
\end{equation}
with $m_*=\min_q\nu/(\sqrt2q)\simeq1.4173$ at $q_*\simeq0.8805$.
\end{proposition}

\begin{proof}[Derivation]
Near the surface, $\Phi-\mu$ is linear with slope $F$ across the layer.
For $q=k\delta=O(1)$ the Poisson response is smaller than the local compressibility
term: $\delta\Phi\sim\delta\rho/k^2$, whereas
$\delta\mu_{\rm loc}\sim\kappa\rho^{-1/3}\delta\rho$.
Their ratio scales as $\rho^{1/3}\delta^2/\kappa=O(\delta^{5/2})$,
which motivates the Cowling approximation in this large-$l$ limit. With $\chi=\chi_*+ue^{iq\tilde y-i\nu\tau}+v^*e^{-iq\tilde y+i\nu\tau}$, $\tilde y$ a tangential coordinate,
$\mathcal A_q=L_q+\tfrac23\chi_*^{4/3}$ and $\mathcal B=\tfrac23\chi_*^{4/3}$, the Bogoliubov equations
$\nu u=\mathcal A_qu+\mathcal Bv$, $-\nu v=\mathcal Bu+\mathcal A_qv$ give Eq.~(\ref{eq:bdg}) for $f=u+v$. Since
$L_0\chi_*=0$ and $\chi_*>0$, $L_q\ge q^2>0$, so $\nu^2>0$. For $q\to0$ the eigenfunction
reduces to the surface gravity wave of the TF layer, $\nu^2=2q$, which in physical
units is $\omega^2=Fk$. Restoring units, $\omega=\nu F\delta/\heff$; with
$F=\RTF^{-2}$ and Eq.~(\ref{eq:delta}), $F\delta^2/(\heff\RTF)=2^{-1/2}\wK(\heff^2/2\RTF)^{1/6}$,
which gives Eqs.~(\ref{eq:wl_layer}) and (\ref{eq:wnuc}). The corrections are of relative
order $\delta/\RTF$ (curvature of the surface and of $\Phi$) and $1/l$.
\end{proof}

Equation~(\ref{eq:wnuc}) gives $\wLandau\propto\heff^{1/3}\propto N^{-1/9}$ and $l_*\propto N^{2/9}$,
while $\wcone\propto N^{-1/3}\ln N$. Hence
\begin{equation}
\frac{\wLandau}{\wcone}\simeq\frac{m_*}{C_{3/2}}\,\frac{\RTF^{1/2}(2\RTF)^{-1/6}\heff^{-2/3}}{\ln(\RTF/\xi_0)+c_1}\to\infty.
\label{eq:ratio}
\end{equation}
Thus the two energy scales separate at large $N$: a central vortex can
have lower rotating-frame energy while the tested surface branch still has
positive excitation energies. The distinction is between a finite vortex-energy difference and a linear
surface criterion. Neither threshold by itself gives a vortex-entry rate or a
nonlinear rotating phase diagram.

\section{Numerical method}
\label{sec:methods}

Ground states are computed by Riemannian preconditioned conjugate-gradient descent
on the unit sphere, with a Sobolev preconditioner $(\heff^2k^2/2+\sigma)^{-1}$, $\sigma$
the median of $\tfrac23\kappa\rho^{2/3}$ over the bulk, Polak--Ribi\`ere updates, a
Barzilai--Borwein trial step and Armijo backtracking on the full energy
\cite{BaoDu2004,BaoCai2013}. The radial and axisymmetric stationary states have relative residuals
$\|(H-\mu)\psi\|/\max(1,|\mu|)<2\times10^{-8}$, at which the line search reaches the round-off
of the energy.

\paragraph*{Radial solver.} The spherical star is computed for the reduced amplitude $r\psi(r)$ on a
cell-centred grid with the odd extension at $r=0$ and second-order differences; the
potential is the exact discrete shell sum $\Phi_i=-\sum_{j<i}m_j/r_i-\sum_{j\ge i}m_j/r_j$,
which is the gradient of a symmetric discrete energy. Each $\heff$ is solved at
$h=\heff/12$ and $\heff/18$ and Richardson-extrapolated.

\paragraph*{Axisymmetric solver.} States $f(s,z)e^{i\ell\varphi}$ are discretized on a
cell-centred $(s,z)$ grid with the conservative five-point operator; the energy is a
discrete quadratic form whose gradient is the discrete Hamiltonian. The preconditioner
and the Poisson solver use fast diagonalization (DST-I in $z$, eigenbasis of the
symmetrized radial operator), with Dirichlet values of $\Phi$ from the multipole
expansion of $\rho$ ($l\le16$; even $l$ only for the $z\to-z$ symmetric stationary
states, all $l$ for the odd surface modes). On a
Gaussian density the Poisson solver has relative error $5.6\times10^{-5}$ at $h=0.01$
and order $2.00$; the static TF star reproduces Eq.~(\ref{eq:TF0}) to $9\times10^{-8}$
in $\mu$. $\wcone$ is computed at $h=\heff/5$ and $\heff/7.5$ and
Richardson-extrapolated. A general $\lambda$ is implemented by the substitution
$\heff\to2\sqrt\lambda\,\heff$, $\ell\to\ell/(2\sqrt\lambda)$ in the kinetic operator,
which leaves the flow energy unchanged. Boxes are chosen so that $\rho<10^{-12}\rho_c$
at the boundary.

\paragraph*{Surface modes.} A velocity kick $\psi\to\psi\exp(i\beta f_l/\heff)$ with
$f_l(\bx)=\RTF^2(r/\RTF)^lP_l(z/r)$, $P_l$ the Legendre polynomial, multiplied by
$\exp[-(r-\RTF)^2/0.2^2]$ for $r>\RTF$, and $\beta=4\times10^{-3}/l$ (equal surface
velocity for every $l$) excites the nodeless degree-$l$ surface response. It is
propagated by Strang splitting with the kinetic step exact in the
fast-diagonalization basis; both substeps are unitary in the discrete norm, which
changes by $4\times10^{-14}$ over 200 kinetic steps of a random state. The time step
is capped so that the largest kinetic phase per step is below 2, above which
split-step propagation develops a resonance instability. Frequencies are
least-squares fits over six periods, with statistical errors below $6\times10^{-5}$
and agreement with the windowed Fourier peak, limited by the record length, to
$2.5\times10^{-3}$. An untapered kick of fixed $\beta$ is nonlinear for $l\gtrsim6$
(at $\heff=0.025$ it shifts $\omega_8$ by $+3.5\%$), so it is used only for $l=2$.
For a harmonic trap without gravity, where $\omega_l=\sqrt l\,\omega_{\rm ho}$ for any
polytropic equation of state, the procedure extrapolated to $\heff\to0$ gives
$\omega_2=1.41376(22)$ and $\omega_4=1.993(3)$, low by $3\times10^{-4}$ and
$3.5\times10^{-3}$ (about twice the quoted uncertainty); we take this as its
accuracy.

\paragraph*{Other checks and constants.} At $\heff=0.1$ the $\ell=0$ and $\ell=1$
states were recomputed on Cartesian $160^3$ and $192^3$ grids (single precision,
spectral kinetic operator, cell-averaged $1/r$ kernel, $\ell=1$ kept centred by
projection on the $C_4$ sector); the 3D value $\wcone=0.271486$ agrees with the
axisymmetric $0.271573$ to $3\times10^{-4}$. Equation~(\ref{eq:TFrot}) is iterated on
the axisymmetric grid inside the equatorial Roche barrier, and $\wshed$ is located
by bisection to $3\times10^{-5}$. The constants $A$, $K$, $M_1$, $c_{\rm geo}$ and
$\gamma(\lambda)$ follow from the Lane--Emden solution and one-dimensional
boundary-value solutions of Eqs.~(\ref{eq:inner}) and (\ref{eq:core}).
Equation~(\ref{eq:bdg}) is discretized with second-order differences on
$-40\le\eta\le12$; minimization of the sparse spectrum gives $m_*=1.417319$ and
$1.417334$ at 1200 and 6000 points, and $1.417334$ on $[-30,10]$, with
$q_*\simeq0.8805$. The algebraic and functional-analytic steps of
Theorem~\ref{thm:energy} not relying on concentration compactness or Sobolev
inequalities, the Bogoliubov reductions, unit conversions and power counting were
checked in Lean~4 with Mathlib.

\paragraph*{Spherical spectrum.} We also solve the radial Bogoliubov problem with
the linearized Poisson response at each integer $l$
(Appendix~\ref{app:spectrum}), on ground states recomputed for each radial grid;
this has no kick-amplitude or time-step error but shares the radial ground-state
solver. Scans over $2\le l\le30$, $32$, and $36$ at $\heff=0.035$, $0.025$, and
$0.0175$, and over eight or more consecutive $l$ around the minimum at five smaller
$\heff$ down to $0.0032$, resolve both sides of the surface minimum; at each $l$ the
lowest positive eigenvalue that is nodeless and surface-localized is retained.
Grids $h=\heff/8$, $\heff/12$, and $\heff/18$ and an enlarged radial box test the
minima.


\section{Results}
\label{sec:results}

\subsection{Static star}
\label{sec:static}

\begin{figure*}[!tbp]
\includegraphics[width=\textwidth]{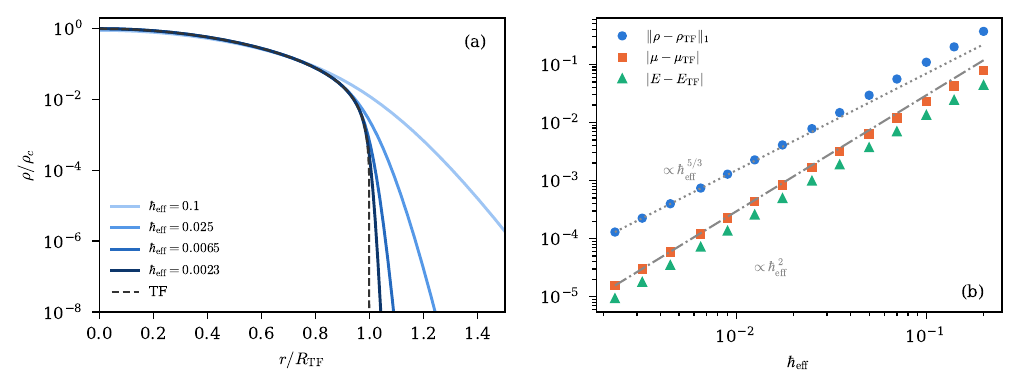}
\caption{Static star. (a) Density profiles for four values of $\heff$ against the
TF profile (Lane--Emden index $3/2$). (b) $L^1$ density deviation and deviations of $\mu$ and $E$ from
TF; guides $\propto\heff^{5/3}$ and $\propto\heff^{2}$.}
\label{fig:static}
\end{figure*}

\begin{figure*}[!tbp]
\includegraphics[width=\textwidth]{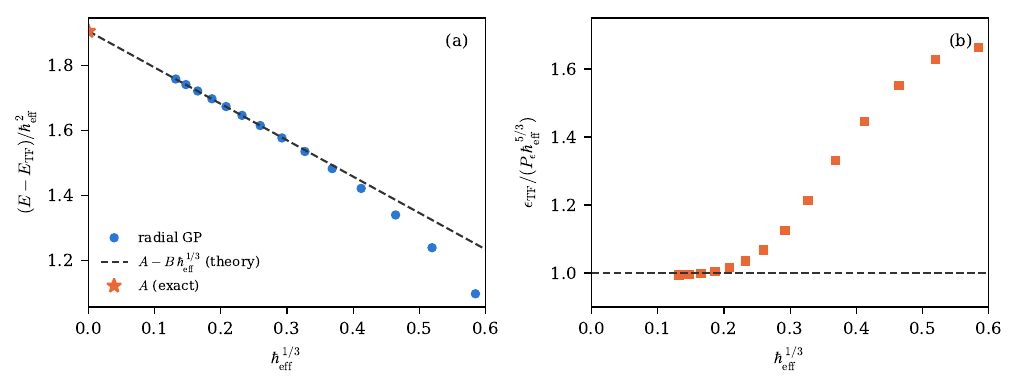}
\caption{Test of Theorem~\ref{thm:energy} and Proposition~\ref{prop:layer}.
(a) $(E-E_{\rm TF})/\heff^2$ against $\heff^{1/3}$; the star is the exact limit $A$,
the dashed line $A-B\heff^{1/3}$ with $B$ from Eq.~(\ref{eq:inner}). (b) $L^1$ density
error $\epsilon_{\rm TF}=\|\rho-\rTF\|_1$ in units of the predicted $P_\epsilon\heff^{5/3}$.}
\label{fig:theory}
\end{figure*}

Table~\ref{tab:static} and Figs.~\ref{fig:static} and \ref{fig:theory} cover
$0.0023\le\heff\le0.2$ ($16\lesssim N\lesssim10^7$), with Richardson errors below
$2.2\times10^{-4}$ of $\epsilon_{\rm TF}=\|\rho-\rTF\|_1$ and virial residuals below
$2.2\times10^{-8}$. Every entry satisfies Theorem~\ref{thm:energy}(i), with
$(E-E_{\rm TF})/(A\heff^2)$ between $0.58$ and $0.92$. The deficit
$A-(E-E_{\rm TF})/\heff^2$ follows a power law with exponent $0.3329\pm0.0002$ over the
five smallest $\heff$, against the predicted $1/3$; with the exponent fixed and a next
term $\propto\heff^{2/3}$, the prefactor is $B=1.114$, against $1.1180$
[Fig.~\ref{fig:theory}(a)]. The chemical potential behaves alike,
$(\mu-\mu_{\rm TF})/\heff^2=2.95$ at $\heff=0.0023$ against $\tfrac53A=3.17$, with a
deficit $\propto\heff^{0.3330\pm0.0010}$.

With $T=\heff^2\mathcal T$, the virial relation $\mu=7E/3-2T/3$, and Hellmann--Feynman,
$T=\heff E'(\heff)/2$, a termwise-differentiable expansion (\ref{eq:Eexp}) predicts
\begin{equation}
T-A\heff^2\simeq-\tfrac76B\heff^{7/3},\quad
\mu-\mu_{\rm TF}-\tfrac53A\heff^2\simeq-\tfrac{14}{9}B\heff^{7/3}.
\label{eq:staticcross}
\end{equation}
At $\heff=0.0125$ and $0.0065$ the extracted coefficients are $1.3039$ and $1.3004$
against $7B/6=1.3044$, $1.7341$ and $1.7343$ against $14B/9=1.7392$, and $1.1158$ and
$1.1146$ against $B$. These are linked by the virial relation, whose residual is below
$6\times10^{-9}$, and are not independent determinations; at the smallest $\heff$
subtraction amplifies discretization errors.

The local exponent $\dd\ln\epsilon_{\rm TF}/\dd\ln\heff$ rises from $1.73$ to $1.93$ near
$\heff=0.03$ and decreases to $1.670$ for the smallest pair, a crossover between a
bulk term $O(\heff^2)$ and the surface term $O(\heff^{5/3})$. The ratio
$\epsilon_{\rm TF}/(P_\epsilon\heff^{5/3})$ [Fig.~\ref{fig:theory}(b)] decreases to $1.005$ at
$\heff=0.0065$ as the bulk contribution, of relative order $\heff^{1/3}$, dies out, and
then falls slightly below unity, to $0.994$ at $\heff=0.0023$. A positive bulk term
cannot produce this undershoot, which exceeds the Richardson error by more than an
order of magnitude; it is consistent with the curvature correction to the planar
layer, of relative order $\delta/\RTF\simeq0.012$, absent from
Proposition~\ref{prop:layer}.

\begin{table*}[!tbp]
\caption{Static star (Richardson-extrapolated radial solver). TF values:
$\mu_{\rm TF}=-0.598975$, $E_{\rm TF}=-0.256704$; $A=1.90474$.}
\label{tab:static}
\begin{ruledtabular}
\begin{tabular}{ccccccc}
$\heff$ & $N$ & $\mu$ & $E$ & $(E-E_{\rm TF})/\heff^2$ & $\epsilon_{\rm TF}$ & $\epsilon_{\rm TF}/(P_\epsilon\heff^{5/3})$\\ \hline
0.2    & 15.6 & $-0.5197412$ & $-0.2128137$ & 1.0972 & $3.71\times10^{-1}$ & 1.663\\
0.1    & 125  & $-0.5757794$ & $-0.2433078$ & 1.3396 & $1.09\times10^{-1}$ & 1.552\\
0.05   & $1.0\times10^3$ & $-0.5926644$ & $-0.2529986$ & 1.4820 & $2.94\times10^{-2}$ & 1.332\\
0.025  & $8.0\times10^3$ & $-0.5973089$ & $-0.2557183$ & 1.5764 & $7.84\times10^{-3}$ & 1.125\\
0.0125 & $6.4\times10^4$ & $-0.5985418$ & $-0.2564464$ & 1.6458 & $2.27\times10^{-3}$ & 1.035\\
0.0065 & $4.6\times10^5$ & $-0.5988545$ & $-0.2566319$ & 1.6967 & $7.41\times10^{-4}$ & 1.005\\
0.0032 & $3.8\times10^6$ & $-0.5989451$ & $-0.2566857$ & 1.7404 & $2.25\times10^{-4}$ & 0.995\\
0.0023 & $1.0\times10^7$ & $-0.5989594$ & $-0.2566943$ & 1.7575 & $1.30\times10^{-4}$ & 0.994\\
\end{tabular}
\end{ruledtabular}
\end{table*}

\subsection{Central-vortex energy}

\begin{figure*}[!tbp]
\includegraphics[width=\textwidth]{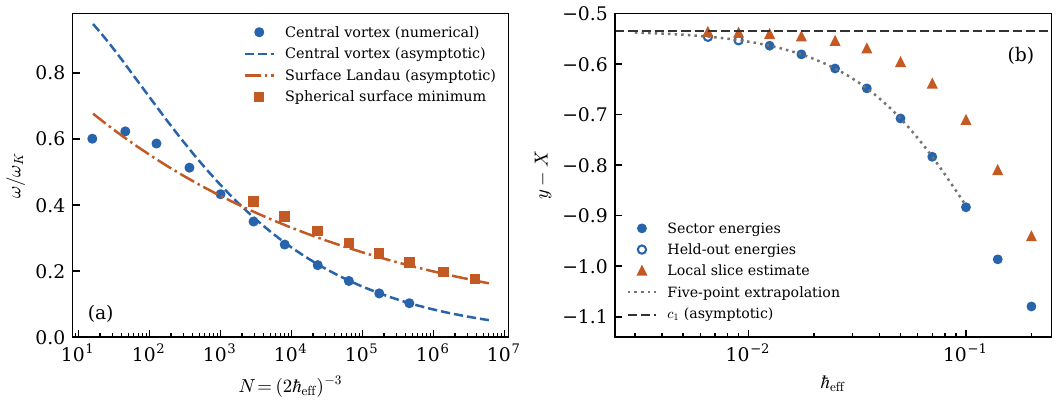}
\caption{(a) Central-vortex energy threshold $\wcone$ (circles) and sampled
surface minima from the spherical spectrum (squares).
Dashed: Eq.~(\ref{eq:wc1}); dash-dotted: the continuum Landau estimate
Eq.~(\ref{eq:wnuc}). Both curves have coefficients fixed by the universal
problems. The spherical scans include every integer multipolarity on both sides of
the minima (Table~\ref{tab:minima}). These are energy scales, not a nonlinear phase diagram. (b) Constant term $y-X$, where $y=\wcone\RTF^2/(C_{3/2}\heff)$ is the
threshold in units of the leading prefactor of Eq.~(\ref{eq:wc1}) and
$X=\ln(\RTF/\xi_0)$; $y-X\to c_1$ as $\heff\to0$. Filled circles: sector energies;
triangles: slice estimate Eq.~(\ref{eq:wloc}) on the computed density; dashed: the theoretical
$c_1=-0.5351$; dotted: a five-point fit of Eq.~(\ref{eq:c1corr}) over
$0.0125\le\heff\le0.05$. Open circles at $\heff=0.009$ and $0.0065$
were computed after the fit and are out of sample.}
\label{fig:wc1}
\end{figure*}

\begin{table*}[!tbp]
\caption{Vortex and surface energy scales. $\wcone$ is computed from
Eq.~(\ref{eq:wc1exact}), with absolute Richardson corrections below
$8\times10^{-5}$. ``Theory'' is the asymptotic central-vortex formula with
$c_1=-0.53509$. The penultimate column is the surface minimum from the spherical spectrum;
the last column is the continuum estimate Eq.~(\ref{eq:wnuc}).
$\wK=0.463568$.}
\label{tab:wc}
\begin{ruledtabular}
\begin{tabular}{cccccccc}
$\heff$ & $N$ & $\wcone$ & $\wcone/\wK$ & Theory & Theory$/\wcone-1$ & $\min_l\omega_l/l$ & $\wLandau$\\ \hline
0.2    & 15.6 & 0.27831 & 0.600 & 0.44048 & $+0.58$ & -- & --\\
0.14   & 45.6 & 0.28860 & 0.623 & 0.38268 & $+0.33$ & -- & --\\
0.1    & 125  & 0.27157 & 0.586 & 0.32344 & $+0.19$ & -- & --\\
0.07   & 364  & 0.23768 & 0.513 & 0.26358 & $+0.11$ & -- & --\\
0.05   & $1.0\times10^3$ & 0.20046 & 0.432 & 0.21332 & $+0.064$ & -- & 0.1980\\
0.035  & $2.9\times10^3$ & 0.16201 & 0.349 & 0.16791 & $+0.036$ & 0.1900 & 0.1758\\
0.025  & $8.0\times10^3$ & 0.12971 & 0.280 & 0.13246 & $+0.021$ & 0.1689 & 0.1571\\
0.0175 & $2.3\times10^4$ & 0.10082 & 0.217 & 0.10202 & $+0.012$ & 0.1487 & 0.1395\\
0.0125 & $6.4\times10^4$ & 0.07859 & 0.170 & 0.07913 & $+0.007$ & 0.1317 & 0.1247\\
0.009  & $1.7\times10^5$ & 0.06113 & 0.132 & 0.06138 & $+0.004$ & 0.1171 & 0.1118\\
0.0065 & $4.6\times10^5$ & 0.04737 & 0.102 & 0.04748 & $+0.002$ & 0.1043 & 0.1003\\
\end{tabular}
\end{ruledtabular}
\end{table*}

\begin{figure*}[!tbp]
\includegraphics[width=\textwidth]{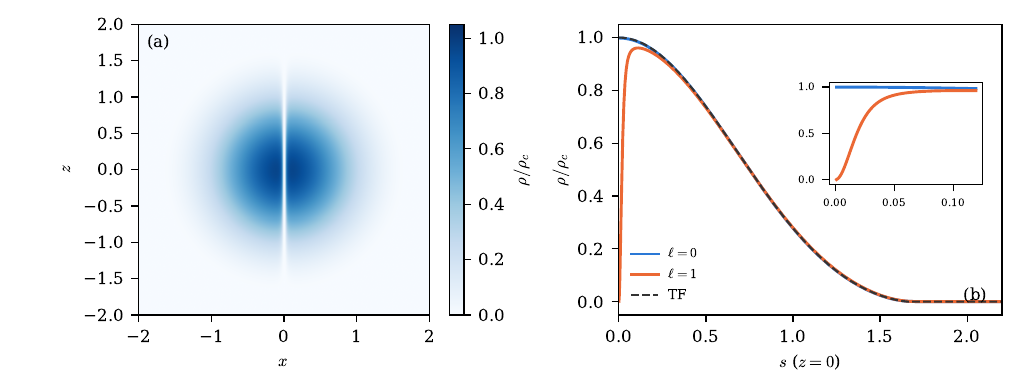}
\caption{Central vortex at $\heff=0.0125$ ($N=6.4\times10^4$). (a) Density of the
$\ell=1$ state in the meridional plane $y=0$. (b) Equatorial profiles of the $\ell=0$ and
$\ell=1$ states against TF as functions of the cylindrical radius $s$; inset: the core,
of width $\sim\xi_0=0.014$. Lengths in units of $r_N$.}
\label{fig:vortex}
\end{figure*}

Table~\ref{tab:wc} compares the exact $\wcone$ of Eq.~(\ref{eq:wc1exact}) with
Proposition~\ref{prop:wc1}, which has no adjustable parameter. The relative
difference decreases monotonically from $19\%$ at $N=125$ to $0.7\%$ at
$N=6.4\times10^4$ and $0.2\%$ at $N=4.6\times10^5$; correspondingly $y-X$ approaches
$c_1$ from below [Fig.~\ref{fig:wc1}(b)]. For $N\lesssim100$ the logarithm is not large
and $\wcone$ even rises with $\heff$, the star being broader than its TF radius. The
vortex state (Fig.~\ref{fig:vortex}) is a straight line vortex through an otherwise
undistorted polytrope. The observed order of convergence of $E_0$ is $2.000$, and the
two-grid Richardson correction to $\wcone$ is at most $2.7\times10^{-4}$ of its value.

\paragraph*{Constant extraction.}
A five-point fit of Eq.~(\ref{eq:c1corr}) over $0.0125\le\heff\le0.05$ gives
$c_1=-0.53372$ and predicts $\wcone=0.061136$ and $0.047371$ at $\heff=0.009$ and
$0.0065$. The subsequently computed values, $0.061134(2)$ and $0.047368(2)$, differ
by $4\times10^{-5}$ and $8\times10^{-5}$ in relative frequency. Including them, fits
over the four to seven smallest $\heff$ give intercepts from $-0.53458$ to $-0.53406$,
moving toward the predicted $-0.53509$ as the window narrows. Adding a term
$\heff^2\ln\heff^{-1}$ or $\heff^2\ln^2\heff^{-1}$ to the ansatz moves the five- to
seven-point intercepts to $[-0.53479,-0.53451]$ and $[-0.53494,-0.53470]$, at the
cost of worse conditioning. Subtracting the slice estimate of Eq.~(\ref{eq:wloc})
leaves a residual whose fitted intercept lies between $2.4\times10^{-4}$ and
$5.9\times10^{-4}$. With finite-grid corrections in $y-X$ of at most $2\times10^{-4}$,
propagated with unrestricted signs (Appendix~\ref{app:vortex}), the extrapolation
sensitivity is $(0.4$--$3.8)\times10^{-3}$, depending on the ansatz; the data
therefore support $c_1$ at the $10^{-3}$ level. A free power fit gives intercepts
between $-0.517$ and $-0.529$; the frequency prediction is more stable than the
limiting intercept.

\paragraph*{Dependence on the gradient coefficient.} Proposition~\ref{prop:wc1} predicts
that $\lambda$ enters $\wcone$ only through $\gamma(\lambda)$, since $c_{\rm geo}$ is a TF
quantity. We repeated the calculation at $\lambda=1/9$ for $\heff=0.05$, $0.025$ and
$0.0125$, obtaining $\wcone=0.191484$, $0.124094$ and $0.075578$. The shift of $y-X$
relative to $\lambda=1/4$ is $-0.1206$, $-0.1509$ and $-0.1620$. The deviations from the
predicted $\gamma(1/9)-\gamma(1/4)=-0.1675$ decrease by factors $2.7$ and $3.0$ per
halving of $\heff$, close to the $2^{5/3}=3.2$ of Eq.~(\ref{eq:c1corr}); extrapolation
with that form gives $-0.1683$ from three points and $-0.1672$ from the two smallest $\heff$
without the logarithm. The limiting shift is consistent with the predicted core-constant difference,
providing a check independent of the geometric constant. At finite $\heff$,
$\lambda$ also changes the background and endpoint corrections.

\subsection{Surface modes and separation of energy scales}
\label{sec:modes}

\begin{figure}[!tbp]
\includegraphics[width=\columnwidth]{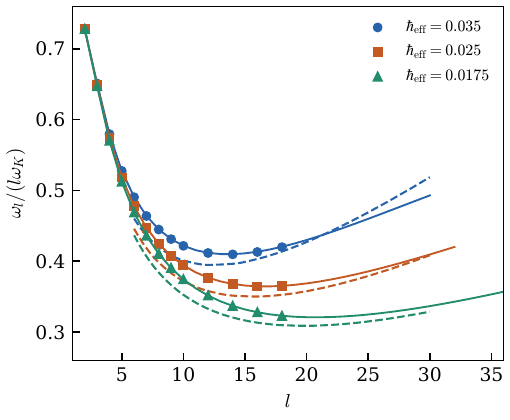}
\caption{Surface-mode energy per unit angular momentum, $\omega_l/l$, in units
of $\wK$ for three values of $\heff$. Symbols are the archived real-time
frequencies; solid lines join the spherical eigenfrequencies at
every integer $l$. Dashed lines are the large-$l$ surface formula
Eq.~(\ref{eq:wl_layer}), shown for $l\ge6$, with no fitted coefficient.}
\label{fig:modes}
\end{figure}

\paragraph*{Quadrupole and time-domain checks.} The $l=2$ frequency extrapolates to
$\omega_2=0.67502(7)=1.456\,\wK$; the quadrupole threshold $\omega_2/2=0.728\,\wK$ lies
well above the high-$l$ surface minima (Fig.~\ref{fig:modes}), so the quadrupole does
not set the surface Landau scale. A quadrupolar stirring instability
\cite{Recati2001} involves a driven, deformed background and is not determined by
this comparison. The $l=2,8$ refinement at $\heff=0.025$ changes the frequencies by
less than $6.1\times10^{-5}$ under grid, time-step, and amplitude variations. The
high-$l$ traces have single-frequency rms residuals of $1.5$--$3.2\%$ of the signal
standard deviation [Fig.~\ref{fig:diagnostics}(a,b)], and early or late fit windows
shift their frequencies by at most $3.6\times10^{-4}$. The spherical
eigenfrequencies agree with the archived $l=10$--$18$ frequencies to
$1.4\times10^{-3}$ or better, identifying them as the fundamental surface response.

\paragraph*{Resolved surface minima.}
The spherical calculation resolves minima at $l=14$, $17$, and $21$ (Table~\ref{tab:minima}
and Fig.~\ref{fig:modes}) and continues the sequence to $l=62$ at $\heff=0.0032$. The
radial density perturbations are nodeless [Fig.~\ref{fig:diagnostics}(c)]; within
$4\delta$ of the TF surface lie more than $99\%$ ($96.8\%$ at the smaller $\heff$) of
their squared density norm and $99.7\%$ ($98.5\%$) of their quasiparticle amplitude
norm. Refining $h=\heff/8$ to $\heff/18$ changes all minima by less than
$5\times10^{-6}$, and enlarging the radial box by $0.5$ changes them by less than
$4\times10^{-8}$.

\begin{table}[!tbp]
\caption{Minima of the localized collective surface band from the spherical
Bogoliubov--Poisson calculation, at $h=\heff/18$. The scan includes all integer
$l$ in the stated range. The continuum asymptotic estimate is
Eq.~(\ref{eq:wnuc}) (Cont.); Rem.\ is the excess of the minimum over it after
removal of the geometric factor $\sqrt{l(l+1)}/l$.}
\label{tab:minima}
\setlength{\tabcolsep}{2pt}
\begin{ruledtabular}
\begin{tabular}{cccccc}
$\heff$ & $l$ range & $l_{\min}$ & $\min_l\omega_l/l$ & Cont. & Rem.\\ \hline
0.035 & 2--30 & 14 & 0.190003 & 0.17579 & 4.42\%\\
0.025 & 2--32 & 17 & 0.168903 & 0.15714 & 4.46\%\\
0.0175 & 2--36 & 21 & 0.148675 & 0.13952 & 4.11\%\\
0.0125 & 23--30 & 26 & 0.131743 & 0.12472 & 3.66\%\\
0.009 & 29--36 & 32 & 0.117105 & 0.11179 & 3.16\%\\
0.0065 & 37--44 & 39 & 0.104281 & 0.10029 & 2.67\%\\
0.0045 & 47--54 & 50 & 0.091578 & 0.08872 & 2.20\%\\
0.0032 & 60--67 & 62 & 0.081266 & 0.07919 & 1.80\%\\
\end{tabular}
\end{ruledtabular}
\end{table}

\paragraph*{Comparison with surface asymptotics.}
The planar formula predicts $l_*\simeq12.3$, $15.4$, and $19.5$ at the three largest
$\heff$; the finite-size minima lie at somewhat larger $l$ and exceed the continuum
estimate by $8.1\%$ at $\heff=0.035$ and $2.6\%$ at $\heff=0.0032$. About half of the
excess is the geometric factor $\sqrt{l(l+1)}/l\simeq1+1/(2l)$ of
Eq.~(\ref{eq:wl_layer}). The remainder (Table~\ref{tab:minima}) stays near $4\%$ for
the three largest $\heff$ and then decreases steadily to $1.80\%$; its local exponent
with respect to $\delta/\RTF$ rises from $0.52$ to $0.88$, the approach expected of a
preasymptotic correction of relative order $\delta/\RTF$. A fit to
$a+b\,\delta/\RTF+c\,(\delta/\RTF)^2$ over the four to six smallest $\heff$ gives
$|a|<5\times10^{-4}$ and $b=1.35$--$1.44$, so the minima extrapolate to
Eq.~(\ref{eq:wnuc}) with $m_*=1.4173$ within $5\times10^{-4}$. The local scaling
exponent of the minima decreases from $0.359$ to $0.350$, toward $1/3$. The Cowling
approximation raises the spherical frequency by only $0.20\%$ at $(\heff,l)=(0.025,16)$
and $0.012\%$ at $(0.0032,62)$.

\paragraph*{Separation of energy scales.}
The resolved surface minima exceed $\wcone$ by factors $1.17$, $1.30$, $1.47$,
$1.68$, $1.92$, and $2.20$ for $0.035\ge\heff\ge0.0065$, where both are computed,
with a ratio growing with $N$ as Eq.~(\ref{eq:ratio}) predicts. This does not exclude
particle escape or establish a nonlinear vortex-entry threshold. At $N=10^3$ the
continuum estimate $\wLandau=0.198$ is too close to $\wcone=0.200$ to resolve their
order from the leading formula.


\section{Discussion}
\label{sec:discussion}

The calculations identify two regions in which a small gradient coefficient has a
leading effect, the surface and the vortex core. The leading static energy
correction is the gradient functional evaluated on the TF profile; relaxation of
the surface lowers it at order $\heff^{7/3}$ and dominates the integrated density
error at order $\heff^{5/3}$. The central-vortex energy is logarithmically larger
than the static gradient energy, with a constant that separates Lane--Emden
geometry from core structure. The surface spectrum provides a third scale,
$\wLandau\propto N^{-1/9}$, which decreases more slowly than
$\wcone\propto N^{-1/3}\ln N$. Agreement of the total energy with TF therefore
does not determine vortex energetics or the surface Landau threshold.

\subsection{Model dependence}

All asymptotic statements concern the specified ETF functional. The coefficients
depend on $\xi_B$ and $\lambda$, while the exponents follow from the $\rho^{5/3}$
energy density and the nonzero surface slope. Static results at another
$\lambda>0$ follow from the rescaling of Sec.~\ref{sec:model}; the vortex constant
requires $\gamma(\lambda)$, and the surface coefficient is computed only for
$\lambda=1/4$. Neither the core nor the surface becomes a controlled microscopic
gradient expansion as $N$ grows: with $k_F=(3\pi^2n)^{1/3}$ and the dimensional
layer width $d_s=r_N\delta$, a point of the layer with $\rho=O(\delta^{3/2})$ has
$k_Fd_s=(3\pi^2)^{1/3}N^{1/3}\rho^{1/3}\delta=O(N^{1/3}\delta^{3/2})=O(1)$, and
likewise $k_Fr_N\xi=O(1)$ in the core. The computed $B$, $P_\epsilon$, $c_1$, and
$m_*$ are thus well-defined coefficients of the functional, but additional
gradient operators and quasiparticle dynamics can change their microscopic values,
consistent with the separate bulk and edge descriptions of
Ref.~\cite{Hellerman2024}. A long-range Fock-type term, estimated as
$E_{\rm Fock}\sim uN^{4/3}/r_N$ against $|E_{\rm Hartree}|\sim uN^2/r_N$, enters at
relative order $N^{-2/3}\sim\heff^2$, the order of the leading gradient
correction, so $A$ is not the complete microscopic correction of a Hamiltonian
with bare $1/r$ attraction. The pair field also does not resolve fermionic
quasiparticles or pair breaking; comparison with a microscopic superfluid
functional \cite{BulgacYu2003} would test these assumptions. Shell effects
\cite{BrackBhaduri1997}, finite effective range, and finite temperature are absent.

\subsection{Vortex entry and rotating equilibria}

The Landau criterion is a property of the vortex-free surface branch: it marks
where a rotating perturbation can couple to negative-energy surface excitations,
but a drive and the subsequent nonlinear dynamics determine vortex entry.

\paragraph*{Rigid-rotation reference.}
If a dense lattice exists and its coarse-grained flow is rigid, the TF density
satisfies
\begin{equation}
\kappa\rho^{2/3}+\Phi-\tfrac12\omega^2s^2=\mu.
\label{eq:TFrot}
\end{equation}
The sequence of connected solutions ends at the equatorial Roche barrier, at
$\wshed=0.24510=0.529\wK$. This is a reference scale for the rigidly rotating TF
branch, not a bound for the vortex-free or single-vortex branch. $\wcone$ crosses
it near $N\simeq285$ (near $N\simeq404$ when rescaled with the rms radius), which
does not prove a minimum particle number for vortex existence; that requires
stability and mass-loss calculations of few-vortex states at fixed intrinsic
angular momentum.

\paragraph*{Landau-level scale and lattices.}
For $\lambda=1/4$ the transverse Landau-level spacing is $2\heff\omega$. Along the
rigidly rotating TF sequence $\muloc(0)=\kappa\rho(0)^{2/3}$ decreases from $0.806$
to $0.648$ at $\omega=0.995\wshed$, so $\muloc(0)/(2\heff\omega)\gtrsim1.3/\heff$
and the dense bulk does not approach a lowest-Landau-level regime
\cite{Ho2001,Schweikhard2004,Cooper2008,NguyenRougerie2022,LeNguyen2026}. A dense
lattice adds an energy density of order $\omega\heff\rho\ln(b/\xi)$ with
$b=(\heff/\omega)^{1/2}$, which can exceed the static gradient correction; its
density response must include the perturbed Hartree potential, normalization,
and the angular-momentum constraint, and is left open.

\paragraph*{Realizations.}
Extending the light-induced $1/r$ attraction \cite{ODell2000} to a paired
two-component Fermi gas would require assessing the accessible coupling,
photon-scattering heating, and losses; we have not established an experimental
parameter window. For Newtonian attraction the model is a formal
nonrelativistic reference, not a description of a neutron star.

\section{Conclusion}

The TF limit of the self-bound ETF superfluid is controlled in energy by a
regular first-order gradient contribution and in density by a singular surface
layer. Matching this layer and a vortex core to the polytropic bulk determines
the next static correction, the constant in the central-vortex energy, and the
surface Landau scale. The numerical calculations confirm the static
coefficients and the vortex constant at the subpercent and $10^{-3}$ levels, and
the surface Landau coefficient $m_*$ to about $5\times10^{-4}$ after a
quadratic extrapolation of the finite-size minima in $\delta/\RTF$; they also expose the finite-size bias
in extracting limiting coefficients. The separation between the vortex-energy threshold and the
surface criterion is predicted by the formal asymptotics and supported in the
six cases in which both scales are computed. Nonlinear
vortex entry and the stability of rotating few-vortex and lattice states require
additional calculations.

\paragraph*{Data and code availability.} The solver, the archived numerical
results, the analysis scripts, and the Lean~4 proof files are available from
the author upon request and will be deposited in a public repository before
publication.

\begin{acknowledgments}
We sincerely thank Dr. Dinh-Thi Nguyen for his academic insights and inspiration throughout the writing of this article. We are also grateful to the 2026 School of Mathematics and Physics (VIASM x HCMUS) for providing an inspiring academic environment that greatly motivated this work.
\end{acknowledgments}

\appendix
\section{Surface mass cancellation and energy subtraction}
\label{app:surface}

The inner equation implies the stress identity
\begin{equation}
\frac{\dd}{\dd\eta}\left[\chi_*'^2-\eta\chi_*^2
-\frac35\chi_*^{10/3}\right]=-\chi_*^2.
\end{equation}
The corresponding TF stress is $\tfrac25(-\eta)_+^{5/2}$, whose derivative
is $-\chi_{\rm TF}^2$. The difference of these stresses tends to zero at both
ends: at $\eta\to-\infty$ it is $O((-\eta)^{-1/2})$ and at
$\eta\to+\infty$ the inner solution decays. Integration therefore gives
\begin{equation}
\int_{\mathbb R}(\chi_*^2-\chi_{\rm TF}^2)\,\dd\eta=0.
\end{equation}
This signed cancellation is compatible with a nonzero absolute mass integral
$M_1$. The leading surface redistribution thus produces the stated $L^1$
error without shifting the bulk normalization at that order.

For the energy, subtract the TF grand-potential density and the TF amplitude
gradient in the overlap region. The remaining surface integral is $K$, with
prefactor
$4\pi\RTF^2F^{5/2}\kappa^{-3/2}\delta^{7/2}$.
Using $\delta=(\heff^2/2F)^{1/3}$ gives the correction
$-B\heff^{7/3}$ in Eq.~(\ref{eq:Eexp}). To make the matching explicit, choose an intermediate distance $a$ with
$\delta\ll a\ll\RTF$. The subtraction includes the TF amplitude gradient
on both sides of this cut; hence its cutoff-dependent contribution is not
counted twice. Because $\chi_*$ and $\chi_{\rm TF}$ share the leading gradient
$\tfrac{9}{16}(-\eta)^{-1/2}$ and the $O((-\eta)^{-1/2})$ terms of the local energy
cancel, the subtracted integrand decays as $(-\eta)^{-7/2}$; the missing
contribution to $K$ is $O((a/\delta)^{-5/2})$, of dimensional size
$O(\delta^{7/2}(\delta/a)^{5/2})=o(\delta^{7/2})$.
Within this matching window the geometric corrections are at most of relative
order $a/\RTF$, which vanishes as the intermediate cut tends to the edge.
In the overlap, the leading density correction behaves as
$\Delta\rho=O(\heff^2(\RTF-r)^{-3/2})$; the quadratic local energy left
outside the cut scales as $O(\heff^4a^{-5/2})=o(\delta^{7/2})$.
These estimates require a cut for which the outer and inner expansions
both hold, as assumed in Proposition~\ref{prop:layer}.

The first variation of the TF local plus Hartree energy is
$\mu_{\rm TF}\int_{\mathbb R^3}\Delta\rho\,\dd\bx$ plus
$\int_{\{|\bx|>\RTF\}}(\Phi_{\rm TF}-\mu_{\rm TF})\Delta\rho\,\dd\bx$, where $\Delta\rho=\rho_{\heff}-\rTF$ and $\Phi_{\rm TF}$ is the Hartree potential of $\rTF$.
At fixed mass the first term vanishes; the exterior linear-potential cost
is retained in $\eta\chi_*^2$ in the inner grand potential, rather than
discarded. The signed layer-mass
identity above removes a normalization shift of order $\delta^{5/2}$.
The remaining nonlocal term is the quadratic Coulomb form
$\tfrac12\int_{\mathbb R^3}\Delta\rho\,\Delta\Phi\,\dd\bx$, with $\nabla^2\Delta\Phi=4\pi\Delta\rho$. For the matched layer and outer tail,
$\|\Delta\rho\|_{6/5}=O(\delta^{7/3})$; the Hardy--Littlewood--Sobolev
bound therefore gives $O(\delta^{14/3})=o(\delta^{7/2})$ for this term.
Smooth $O(\heff^2)$ bulk normalization changes are also beyond the displayed
surface energy after the first variation is removed. These are formal
power-counting estimates for the matched profile, not uniform error bounds
for exact minimizers; they determine the coefficient without proving the
stated asymptotic remainder.

\section{Vortex energy in a nonuniform column}
\label{app:vortex}

For the axisymmetric $\ell=1$ state the flow energy is
$\tfrac12\heff^2\int_{\mathbb R^3}\rho\,s^{-2}\,\dd\bx$ exactly. Split each slice at $\xi\ll a\ll R_\perp(z)$.
Away from the endpoints, for $s>a$ the leading profile is $\rTF$ and
$\int_a^{R_\perp}\rTF(s,z)\,\dd s/s=\rTF(0,z)[\ln(R_\perp/a)+J(z)]$ with
$J(z)=\int_0^{R_\perp}[\rTF(s,z)/\rTF(0,z)-1]\,\dd s/s$, convergent since
$\rTF(s,z)-\rTF(0,z)\propto s^2$. For $s<a$ the density is locally uniform and the
excess grand potential, including the amplitude and local-energy changes, is
$\pi\rTF(0,z)\heff^2[\ln(a/\xi(z))+\gamma(\lambda)]$ by Eq.~(\ref{eq:gamma}). The
dependence on $a$ cancels, and summing over slices,
\begin{align}
E_1-E_0={}&\pi\heff^2\int_{-\RTF}^{\RTF}\rTF(0,z)\nonumber\\
&\times\Big[\ln\frac{R_\perp(z)}{\xi(z)}+J(z)+\gamma\Big]\dd z+o(\heff^2).
\end{align}
The use of a grand-potential excess in each core is essential at fixed
mass. Let $\Delta M<0$ be the missing mass of the matched vortex profile
before normalization. Locally the core excess is
$\Delta E-\mu_{\rm TF}\Delta M$; restoring $-\Delta M$ in the bulk changes
the energy by $-\mu_{\rm TF}\Delta M$ to first order. Thus the displayed
grand-potential excess is precisely the fixed-mass energy excess at that
order. Away from the line ends, the $s^{-2}$ density tail gives
$\Delta M=O(\heff^2\ln\heff^{-1})$. Assuming a bounded bulk susceptibility
after fixing translations, its residual normalization cost is quadratic,
$O(\heff^4\ln^2\heff^{-1})=o(\heff^2)$. Core nonlinearities are already
included in $\gamma$; this estimate concerns the additional smooth bulk
readjustment. Endpoint errors remain subject to the matching assumptions
of Proposition~\ref{prop:wc1}.

With $\xi(z)/\xi_0=[\rTF(0,z)/\rho_c]^{-1/3}$ and
$\pi\int_{-\RTF}^{\RTF}\rTF(0,z)\,\dd z=C_{3/2}/\RTF^2$, this is Eq.~(\ref{eq:wc1}) with
Eq.~(\ref{eq:cgeo}). The scaling estimate at the ends of the line motivates a correction
of order $\heff^{5/3}\ln\heff^{-1}$ in $y-X$
[Eq.~(\ref{eq:c1corr})].

The extrapolation sensitivity quoted in Sec.~\ref{sec:results} writes the fitted
intercept as $\hat c_1=\sum_i w_i(y_i-X_i)$, with columns normalized before the
least-squares solve, and propagates the finite-grid corrections $e_i$ as
$\sum_i|w_i|e_i$. This unrestricted-sign envelope assesses conditioning; it is not a
bound on all numerical errors.

\section{Spherical spectrum and signal diagnostics}
\label{app:spectrum}

\begin{figure*}[!tbp]
\includegraphics[width=\textwidth]{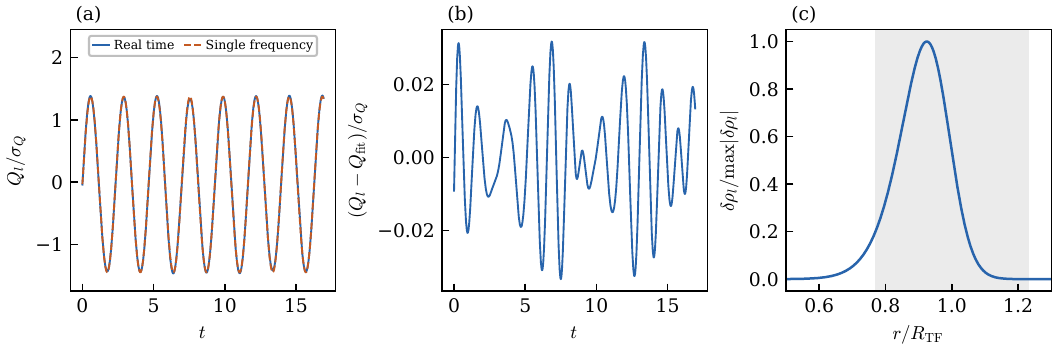}
\caption{Surface-mode diagnostics. (a) Archived real-time signal
$Q_l(t)=\int_{\mathbb R^3}\rho(\bx,t)f_l(\bx)\,\dd\bx$ (mean removed) at
$\heff=0.025$, $l=16$, with $f_l$ the tapered kick profile of Sec.~\ref{sec:methods},
in units of its standard deviation $\sigma_Q$, and the fitted
sinusoid with a linear trend, $Q_{\rm fit}$. (b) The normalized residual
$(Q_l-Q_{\rm fit})/\sigma_Q$. Time $t$ in units of $\Omega_N^{-1}$.
(c) Radial density perturbation $\delta\rho_l(r)$ from the spherical eigenproblem
at the same $\heff,l$, normalized to its maximum magnitude. The shaded band
is $|r-\RTF|<4\delta$.}
\label{fig:diagnostics}
\end{figure*}

Let $p(r)=\sqrt{\rho(r)}$ be the radial stationary amplitude. With a spherical
harmonic of degree $l$, define $f_\pm=r(u\pm v)$ and $P=r\,\delta\Phi$. Linearization gives
\begin{align}
\heff\omega_lf_+&=L_lf_-,\nonumber\\
\heff\omega_lf_-&=\left(L_l+\tfrac43\kappa p^{4/3}\right)f_++2pP,\label{eq:radialbdg}\\
\left[\partial_r^2-\frac{l(l+1)}{r^2}\right]P&=4\pi pf_+,\nonumber
\end{align}
where
\begin{equation}
L_l=-\frac{\heff^2}{2}\partial_r^2+
\frac{\heff^2l(l+1)}{2r^2}+\kappa p^{4/3}+\Phi-\mu.
\end{equation}
Regular solutions behave as $r^{l+1}$ at the origin. The amplitude perturbations
vanish at the outer numerical boundary $r=r_{\max}$; the potential is matched to the exterior
multipole, $P\propto r^{-l}$. Centered second-order differences on the radial
grid and a sparse shift-invert eigenvalue solve determine the lowest positive
frequency. The coupled Poisson field is retained in the eigenproblem; setting
$P=0$ provides the Cowling comparison.

The spherical ground states are recomputed for each grid, with residuals below
$10^{-8}$ in the convention of Sec.~\ref{sec:methods}; at the minima the relative
eigenvalue-equation residual is below $3\times10^{-8}$ for the three largest $\heff$
and below $2.2\times10^{-7}$ otherwise. At $\heff=0.035$, $0.025$, and $0.0175$ the
minima change from $h=\heff/8$ to $\heff/18$ by $3.7\times10^{-7}$, $2.7\times10^{-6}$,
and $3.0\times10^{-6}$ in relative terms. The radial eigenfunctions are nodeless where
their amplitude exceeds $10^{-3}$ of its maximum. Localization is measured with the
norm $\int_0^{r_{\max}}p^2f_+^2\,\dd r$, proportional to
$\int_{\mathbb R^3}|\delta\rho|^2\,\dd\bx$ for a normalized spherical harmonic, and,
so as not to suppress an exterior component through the small background density,
with $\int_0^{r_{\max}}(f_+^2+f_-^2)\,\dd r$, proportional to the positive amplitude norm
$\int_{\mathbb R^3}(|u|^2+|v|^2)\,\dd\bx$. Together with the nodeless structure and the
scan through both sides of each minimum, these tests identify the localized branch
in the calculated range; they do not classify all excitations at unbounded angular
momentum. As a symmetry check, the numerical frequency of the translational mode at
$\heff=0.025$ decreases from $2.08\times10^{-3}\wK$ to $8.72\times10^{-4}\wK$ as $h$
decreases from $\heff/8$ to $\heff/18$.

For the archived time traces at $(\heff,l)=(0.035,14)$, $(0.025,16)$, and
$(0.0175,18)$, refitting a sinusoid plus a linear trend on the early or late two
thirds of the window shifts the frequency by $2.9$, $3.3$, and $3.6\times10^{-4}$;
the full-window rms residuals, $3.2\%$, $1.5\%$, and $2.4\%$ of the signal standard
deviation, allow weak additional components, so the mode is identified by the
spherical calculation rather than by the one-frequency fit alone.

\bibliography{fermion_rot_tf}

\end{document}